\RequirePackage{fix-cm}
\documentclass{article}

\usepackage{graphicx}
\usepackage{epstopdf}
\usepackage{amssymb}
\usepackage{textcomp}
\usepackage{marvosym}
\usepackage{authblk}
\usepackage{natbib}
\usepackage{fullpage}
\begin{document}

\title{Non-linear dependence and teleconnections in climate data: sources, relevance, nonstationarity}

\author[1]{Jaroslav Hlinka}
\author[1]{David Hartman}
\author[1]{Martin Vejmelka}
\author[2]{Dagmar Novotn\'{a}}
\author[1]{Milan Palu\v{s}}
\affil[1]{Institute of Computer Science\\
	      Academy of Sciences of the Czech Republic} 
\affil[2]{Institute of Atmospheric Physics\\
	      Academy of Sciences of the Czech Republic}

\maketitle

\begin{abstract}

Quantification of relations between measured variables of interest by statistical measures of dependence is a common step in analysis of climate data. The term ``connectivity'' is used in the network context including the study of complex coupled dynamical systems. The choice of dependence measure is key for the results of the subsequent analysis and interpretation. The use of linear Pearson's correlation coefficient is widespread and convenient. On the other side, as the climate is widely acknowledged to be a nonlinear system, nonlinear connectivity quantification methods, such as those based on information-theoretical concepts, are increasingly used for this purpose.

In this paper we outline an approach that enables well informed choice of connectivity method for a given type of data, improving the subsequent interpretation of the results. The presented multi-step approach includes statistical testing, quantification of the specific non-linear contribution to the interaction information, localization of nodes with strongest nonlinear contribution and assessment of the role of specific temporal patterns, including signal nonstationarities. 
In detail we study the consequences of the choice of a general nonlinear connectivity measure, namely mutual information, focusing on its relevance and potential alterations in the discovered dependence structure. 

We document the method by applying it on monthly mean temperature data from the NCEP/NCAR reanalysis dataset as well as the ERA dataset. We have been able to identify main sources of observed non-linearity in inter-node couplings. Detailed analysis suggested an important role of several sources of nonstationarity within the climate data. The quantitative role of genuine nonlinear coupling at this scale has proven to be almost negligible, providing quantitative support for the use of linear methods for this type of data.

\end{abstract}

\section{Introduction}
\label{sec:intro}

In climate dynamics research, analysis of time series data has a central position. 
Detection and quantification of dependence between measured or modeled variables is often of interest. 

Apart from the dependences among different physical quantities, in many contexts the task is given to assess the dependence among the measurements of the same physical variable (i.e. surface air temperature, SAT) measured at many different geographical locations. 

The motivation for such a procedure commonly stems from the need to reduce the dimensionality of high-dimensional original data, such as in the application of Empirical Orthogonal Function analysis ~\citep{Hannachi2007} to uncover the basic modes of dynamics of climate system. 
On the other side, dependence quantification might be used to uncover the complex structure of the climate system using approaches such as graph theory ~\citep{Tsonis2004}.
Other applications exist including those combining the above named, see e.g. ~\citep{Tsonis2011}.

There is a wide range of methods available for detection of dependence between variables. The most widely known and used is Pearson's correlation coefficient, a measure particularly sensitive to linear dependence. 
While the Pearson's correlation detects dependence reliably in the case of multivariate Gaussian probability distributions, it may be suboptimal in the case of complex non-Gaussian dependence patterns. For particular dependence patterns (or bivariate probability distributions), it may also fail to detect statistical dependence between variables of interest completely. 

Note that in the following we use the terms `linear' and `Gaussian' dependence interchangeably to denote patterns of dependence corresponding to bivariate normal distribution. While the latter term is more precise, the former is more commonly used in the general community together with the distinction between linear and nonlinear methods.  


However, alternative measures exist that are able to better reflect potential non-linear dependences. These include the Spearman's ordinal correlation coefficient~\citep{Spearman1904} and Kendal's tau~\citep{Kendall1938}, that are designed to be sensitive to any monotonous dependence pattern, without the restriction to linear relationships. An ultimate alternative to the Pearson's correlation coefficient then lies in the utilization of mutual information~\citep{Shannon1948}, an information-theory based measure that is in principle sensitive to any dependence between variables. For this generality, mutual information is widely used to quantify statistical dependence in complex systems, and has been also introduced to the analysis of climate time series ~\citep{Palus2004, Diks2000}.

As the climatic system is highly nonlinear, it seems well motivated to use nonlinear dependence measures for analysis of the measured time series, as suggested e.g. in ~\citep{Donges2009}. This may in theory allow more sensitive detection and quantification of dependences, potentially uncovering new climatic phenomena. 
On the other side, nonlinear measures such as mutual information may have downsides including more difficult implementation and interpretation, increased computational demands and numerical stability issues.

These considerations motivate the central question of the current report: Does the non-linear component of the climate time series dependences sufficiently motivate the use of nonlinear dependence measures?

It is important to note, that the answer to this question might be complex, and certainly would be domain specific. To deal with this complication, we outline here first a generally applicable framework, and then show the results obtained by analyzing in detail a specific dataset of particular interest. This is the monthly SAT data from the NCEP/NCAR reanalysis dataset ~\citep{Kistler2001,Kalnay1996}, as well as concatenated ERA-40~\citep{Uppala2005} and ERA-INTERIM~\citep{Dee2011} data.
Note that this data has been analyzed in many recent studies, utilizing both linear and nonlinear methods, and so constitutes a well motivated timely and relevant example of application of this framework to guide the decision regarding the method choice.

\section{Materials and Methods}
\label{sec:mat}

Data from the NCEP/NCAR reanalysis dataset~\citep{Kistler2001} have been used. In particular, we utilize the time series $x_i(t)$ of the monthly mean SAT from January 1948 to December 2007 ($T=720$ time points), sampled at latitudes $\lambda_i$ and longitude $\phi_i$ forming a regular grid with a step of $\Delta\lambda=\Delta\phi=2.5^{\circ}$. The points located at the globe poles have been removed, giving a total of $N=10224$ spatial sampling points. 

To explore the generalizability of the results, analogous analysis has been carried out for the ERA-40 dataset~\citep{Uppala2005} concatenated with the ERA-INTERIM dataset ~\citep{Dee2011} (further referred to together as the ERA data).

To minimize the bias introduced by periodic changes in the solar input, the mean annual cycle is removed from the data to produce so-called anomaly time series.

We discuss two dependence measures throughout the paper, Pearson's correlation coefficient (linear correlation) and (nonlinear) mutual information. 

Given two real random variables $X, Y$, the well-known Pearson's correlation coefficient is defined as $$\rho_{X,Y}=\sqrt{\frac{E[(X-E(X))(Y-E(Y)]}{E((X-E(X))^2)E((Y-E(Y))^2)}},$$ where $E()$ denotes the expected value operator. The corresponding finite sample estimate is denoted by $r$.

For two discrete random variables $X, Y$ with sets of values $\Xi$ and $\Upsilon$, the mutual information is defined as $$I(X, Y) = \sum_{x\in\Xi}  \sum_{y\in\Upsilon} p(x,y) \log\frac{p(x,y)}{p(x)p(y)},$$ where $p(x)=Pr \{ X= x\},x\in \Xi$ is the probability distribution function of $X$,  $p(y)=Pr \{ Y= y\},y\in \Upsilon$ is the probability distribution function of $X$ and $p(x,y)=Pr \{ (X,Y)= (x,y)\},x\in \Xi, y \in \Upsilon$ is the joint probability distribution function of $X$ and $Y$. For continuous variables, mutual information is defined by the respective integral. However, in practice the mutual information is estimated using discretization of the theoretically continuous variables.  

When the discrete variables $X, Y$ are obtained from continuous variables on a continuous probability space, then the mutual information depends on a partition $\xi$ chosen to discretize the space. A common choice is a simple box-counting algorithm based on marginal equiquantization method \citep{Palus1993}, i.e., a partition is generated adaptively in one dimension (for each variable) so that the marginal bins become equiprobable. This means that there is approximately the same number of data points in each marginal bin. In this paper we use a simple pragmatic choice of $Q=8$ bins for each marginal variable \citep{Palus2007a}.

Mutual information is a non-negative quantity, with $I(X,Y)=0$ corresponding to independence of the variables $X$ and $Y$, and units depending on the base of the logarithm (base 2 corresponds to 'bits', while natural logarithm with base $e$ corresponds to 'nats', used here).
As the estimation of mutual information from finite sample size is prone to sample size dependent bias, to allow quantitative comparison we carry out an approximate correction by recalibration procedure described elsewhere ~\citep{Hlinka2011Neuroimage}. This procedure is in general based on comparison with samples of the same size and coming from populations with analytically established mutual information.

To elucidate our nonlinearity assessment strategy, we first point out that for a bivariate Gaussian distribution, the correlation $\rho(X,Y)$ of the variables $X,Y$ uniquely defines the mutual information between them, which is given by $I(X,Y) = I_G(\rho(X,Y))\equiv -\frac12\log(1-\rho^2(X,Y))$. However, for a general non-Gaussian bivariate distribution, this equation may not hold. Two cases of bivariate non-Gaussianity can be distinguished.

Firstly, the `simpler' nonlinearity consists in non-linear rescaling of one or both of the variables. Such a rescaling does not affect the mutual information between the variables, however, the correlation may change substantially. Rescaling of this type can be suspected in data e.g. due to non-linear properties of the measurement scale, and may be considered as a bias in the correlation estimation. A remedy commonly adopted is the use of Spearman rank correlation coefficient. Alternative procedure lies in preprocessing the data by applying a monotonous transformation to each variable separately that would render it Gaussian (``marginal normalization''), and computing the correlation on the transformed data. 

A second, more `substantial' type of non-Gaussianity lies in that some bivariate distributions differ from bivariate Gaussian not only in their marginal distributions, but also in the form of the interdependence, which cannot be resolved by univariate rescaling. This substantial non-Gaussianity is the key motivation for the use on nonlinear dependence measures, as the dependence pattern can not be recovered by only considering ranks or other rescaled version of the variables. 

Recently, a quantification method for such deviation from Gaussian dependence has been proposed~\citep{Hlinka2011Neuroimage}, building on the fact that for univariately Gaussian random variables $X,Y$, the correlation gives a lower bound on the mutual information by $$I(X,Y) \geq I_G(\rho(X,Y)),$$ with the minimum obtained for bivariate Gaussian distribution. In particular, one can define the neglected (`extra-normal' or `non-Gaussian') information $$I_E(X,Y)=I(X,Y)-I_G(X,Y)\geq 0.$$

Our investigation of the impact of nonlinear contributions to the climate dependence network is carried out in several steps with increasing level of detail.

As a first step, for each pair of local time series $x_i, x_j; \quad i,j\in{1,...,N}$, the correlation $r(x_i, x_j)$ and mutual information $I(x_i,x_j)$ of each pair of local time series is computed, and their overall relation shown in a scatter plot for visual inspection of systematic deviation from the relation valid under Gaussianity. Minor noisy deviations are of course expected due to estimation of the quantities from finite size samples. The step is repeated for univariately normalized variables to isolate the effect of simple rescaling; the normalized variables are further used in the subsequent analysis. 

Notably, the deviation of the relation among the two estimated quantities $r(x_i, x_j)$ and $I(x_i,x_j)$ from the theoretical prediction under Gaussianity ($I(x_i, x_j)=I_E(r(x_i,x_j))$) could be still attributed to the nonlinear properties of the bi-variate dependences, but also to different estimator properties of correlation and mutual information (including residual mutual information estimator bias uncorrected by the procedure mentioned above). 

To isolate the genuine nonlinearity from apparent differences due to different estimator properties, and to allow more robust quantitative comparison, the mutual information estimates from data $X$ are further compared to mutual information estimates obtained for the respective pairs of \emph{linear surrogate data} $S(X)$. The surrogate data conserve the linear structure (covariance and autocovariance) and hence also the correlations of the original data, but remove non-Gaussianity (non-linearity) from the multivariate distribution. Thus, in the large sample size limit, the mutual information between pairs of variables in the surrogate data should be $$I(S_i,S_j)=I_G(\rho(S_i,S_j))=I_G(\rho(X_i,X_j))\leq I(X_i,X_j).$$

Technically, linear surrogate data are conveniently constructed as multivariate Fourier transform (FT) surrogates \citep{Prichard1994,Palus1997}; i.e. obtained by computing the Fourier transform of the series, keeping unchanged the magnitudes of the Fourier coefficients (the amplitude spectrum), but adding the same random number to the phases of coefficients of the same frequency bin; the inverse FT into the time domain is then performed. 

Thus, instead of comparing the estimators of two different quantities $r(x_i, x_j)$ and $I(x_i,x_j)$ (or using the rescaled version of the first one, i.e. $I_E(r(x_i, x_j))$, as an estimate of linear information based on correlation estimator $r$), we can compare the values $I(x_i,x_j)$ and $I(s_i,s_j)$ obtained using the same estimator on both the original dataset $X$ and its linearized (surrogate) version $S$.

As an added value, generation of the surrogates allows direct statistical testing of the Gaussianity of the studied process, as they provide random samples with predefined covariance structure. For this second purpose, a set of $N_{surr}=99$ such surrogate datasets is generated.


For each pair of locations $i,j\in {1,...,N}$, one can test the hypothesis that the time series $x_i, x_j$ come from a bivariate linear stochastic process with Gaussian dependence among the variables by comparing the obtained mutual information estimate $I(x_i,x_j)$ with the empirical distribution of Gaussian mutual information estimate $I(s_{i,k},s_{j,k}); k=1\ldots, N_{surr}$. The one-sided hypothesis is rejected at significance level $\alpha=0.05$ if the data mutual information is higher than at least $95$ (out of $99$) of the surrogate mutual information values.

Apart from testing statistical significance of the deviations from linearity, we also describe the strength and localization of the effect. This can be conveniently vizualized by estimating for each location the average mutual information between the location $i$ and all other locations: for original data  by $I(i)=\sum_j I(x_i,x_j)/N$ and for a surrogate dataset by $I_{surr}(i)=\sum_j I(s_{i},s_{j})/N$. Geographical rendering of the difference $I_E(i)=I(i)-I_{surr}(i)$ and the relative difference $I_E(i)/I(i)$ allows effective visual inspection of the most substantial nonlinearity localization.

Such automatic localization of substantially nonlinear dependencies can be followed by inspection of the corresponding temporal patterns and expert assessment of their relevance for the studied phenomena. Based on this investigation, the use of nonlinear measures for (a subset of) the data may be recommended to exploit the nonlinear information in the data, or alternatively further postprocessing may be proposed to clean the data from spurious sources of apparent nonlinearity, as shown below.  


In particular, the intermediate results of the investigation motivated the introduction of two additional data preprocessing steps adopted in later sections of the paper. The first is the removal of seasonality in variance (variance normalization), which removes the differences in local temperature variability in different parts of year. To achieve that, standard deviation of temperature anomalies is computed for each month separately and the anomaly data from given month are divided by this standard deviation.
The second additional preprocessing step was the removal of slow trends from the data. For simplicity, only a linear trend was considered here.

\section{Results}
\label{sec:res}

\begin{figure}
\includegraphics[width=\columnwidth]{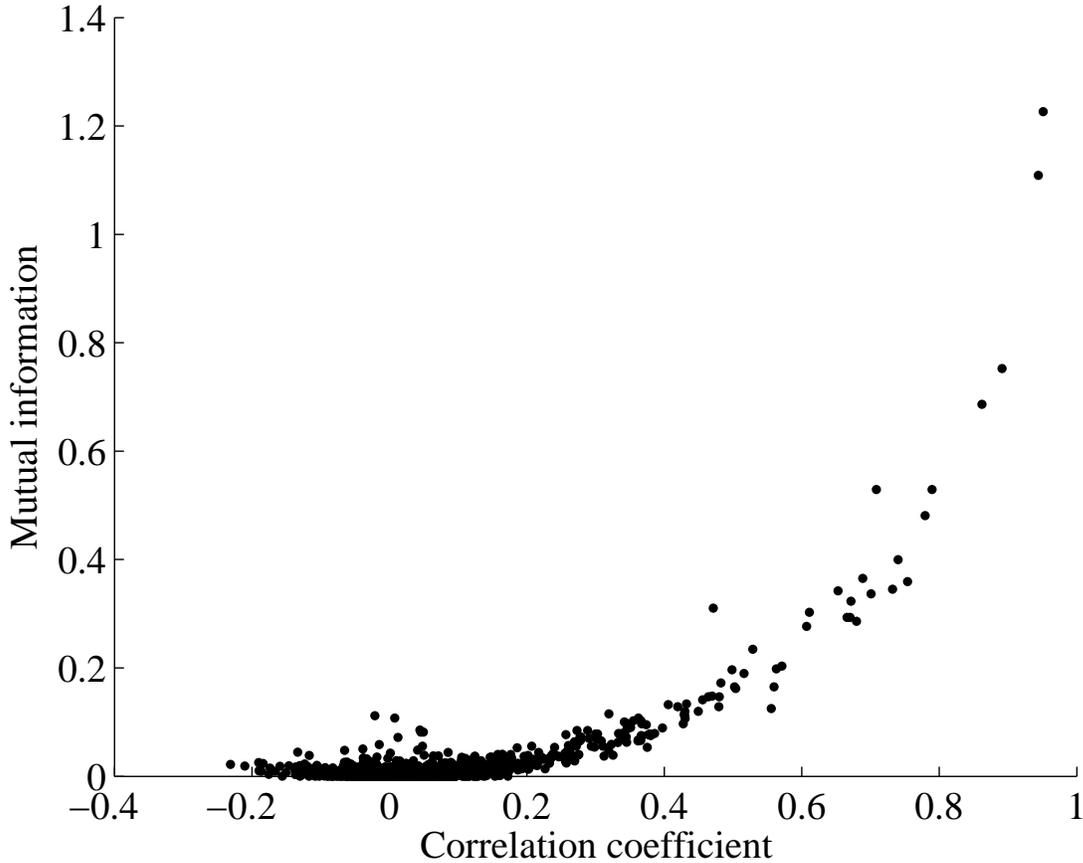} 
\caption{Relation between mutual information and correlation estimates. Each dot corresponds to one time series pair. For visualization purposes, a random selection of 1000 (out of over 5 million) time series pairs shown.}
\label{fig:FigureMIvsC} 
\end{figure}
\begin{figure}
\includegraphics[width=\columnwidth]{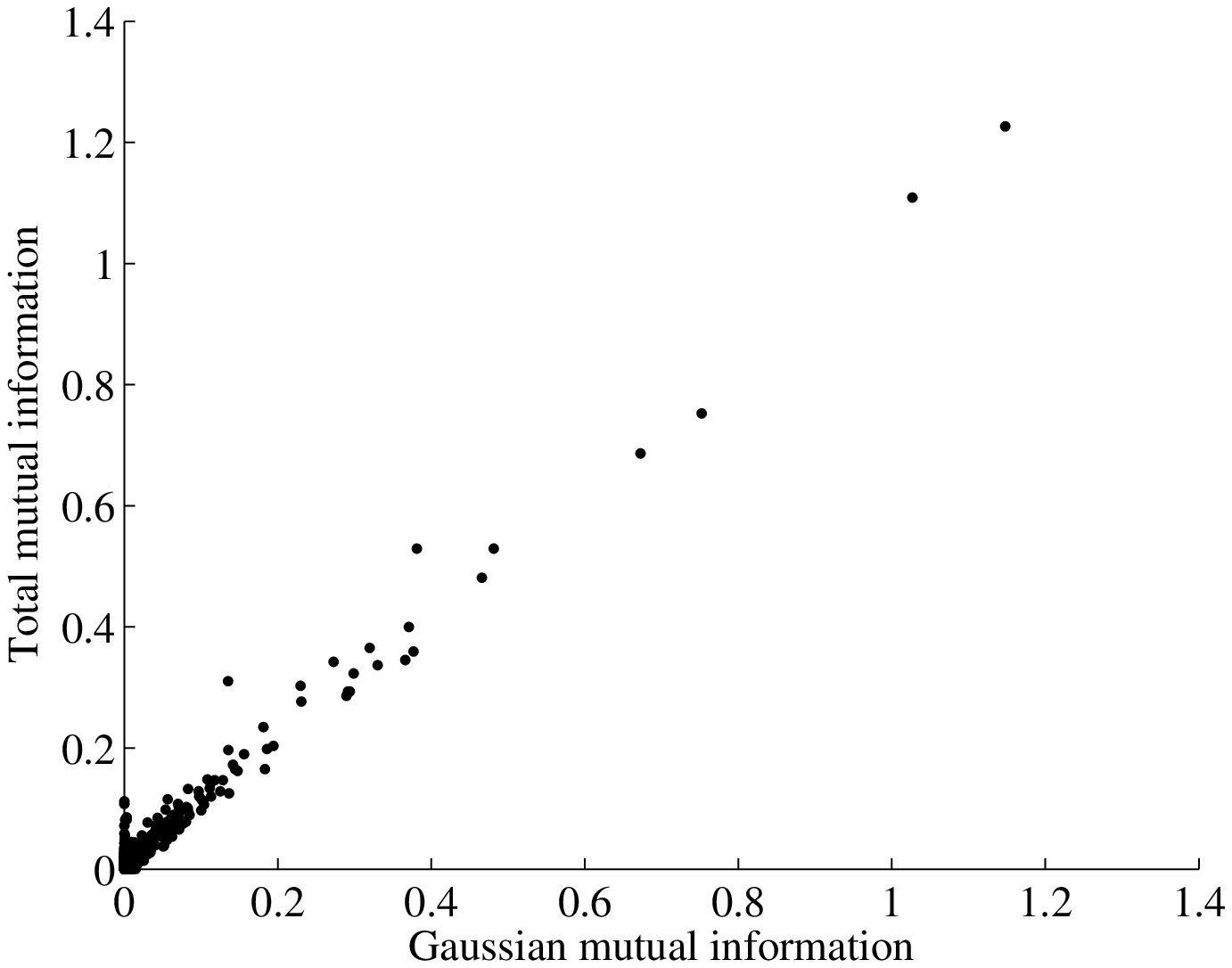} 
\caption{Relation between total and linear mutual information estimates. Each dot corresponds to one time series pair. For visualization purposes, a random selection of 1000 (out of over 5 million) time series pairs shown.}
\label{fig:FigureMIvsMI} 
\end{figure}
\begin{figure}
\includegraphics[width=\columnwidth]{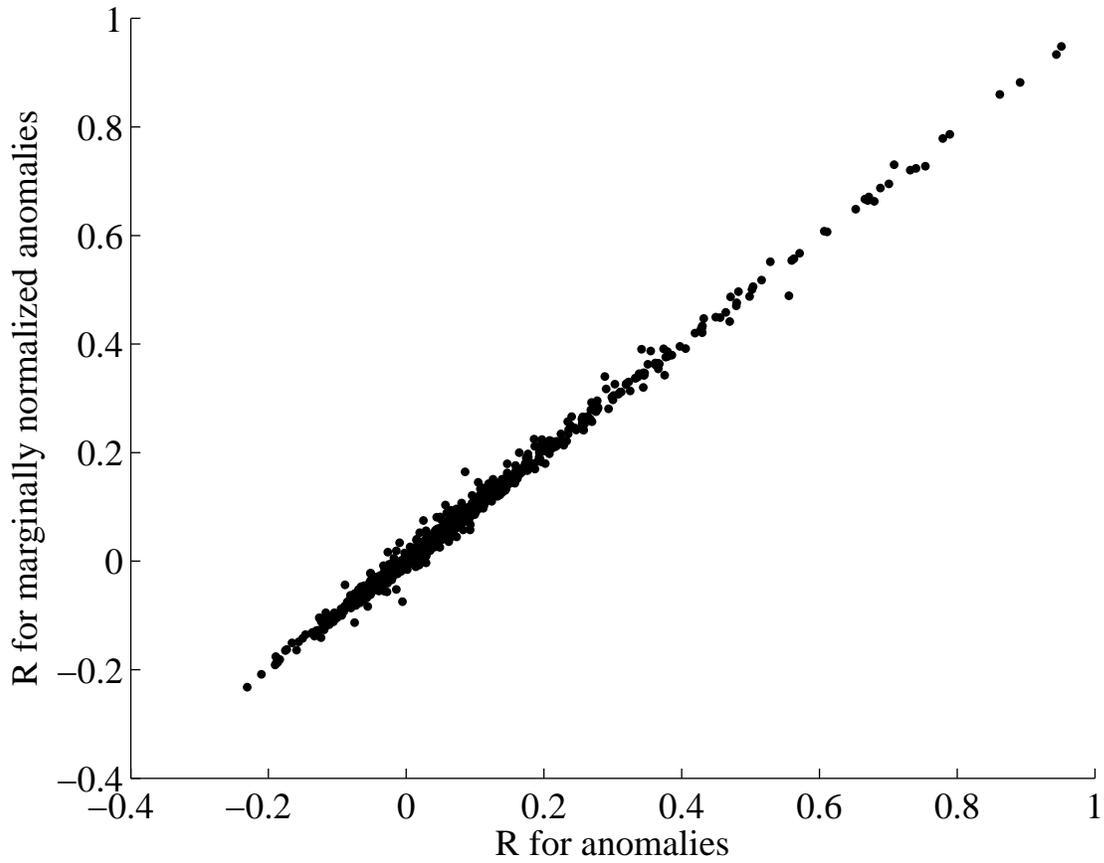} 
\caption{Relation between correlation in original and marginally normalized data. Each dot corresponds to MI of one time series pair. For visualization purposes, a random selection of 1000 (out of over 5 million) time series pairs shown.}
\label{fig:FigureCvsCmarg} 
\end{figure}
\begin{figure}
\includegraphics[width=\columnwidth]{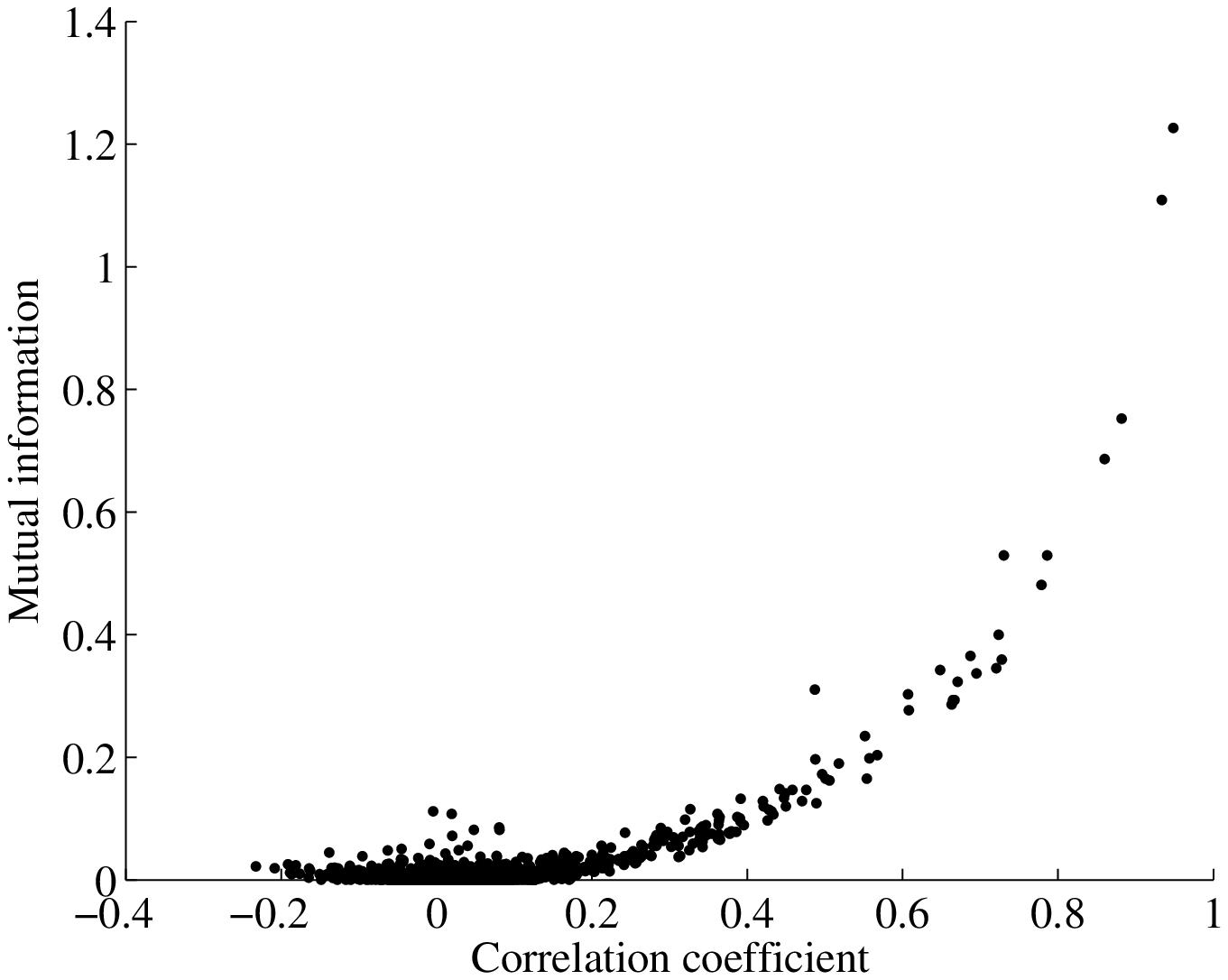} 
\caption{Relation between total and linear correlation estimates in marginally normalized data. Each dot corresponds to MI of one time series pair. For visualization purposes, a random selection of 1000 (out of over 5 million) time series pairs shown.}
\label{fig:FigureMIvsMImarg} 
\end{figure}

The general relation between the mutual information and linear correlation for pairs of SAT anomalies is visualized in Figure~\ref{fig:FigureMIvsC}. The mutual information is in general very strongly related to the correlation. The relation is even clearer after removing the estimator differences by representing the linear dependence in terms of mutual information in the surrogates, see Figure~\ref{fig:FigureMIvsMI}. 
The univariate Gaussianization does not change the correlation coefficients substantially (see Figure~\ref{fig:FigureCvsCmarg}) and does not affect much the relation between data and linear surrogates (see Figure~\ref{fig:FigureMIvsMImarg}).


The deviation from purely Gaussian structure of the data has been tested separately for each pair of variables by comparison with a surrogate dataset distribution.
More than $16$\% time series pairs showed significant at the $p=0.05$ level.

\begin{figure*}
\includegraphics[width=\textwidth]{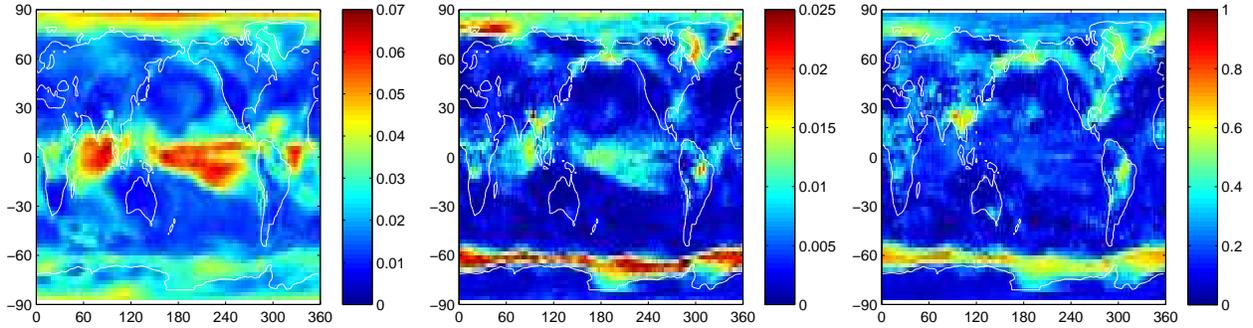}
\caption{Average contribution of nonlinear dependences. Left: average mutual information for each location. Middle: average nonlinear contribution to mutual information $I_E$. Right: average nonlinear contribution relative to total mutual information ($I_E/I$).}
\label{fig:lokalizace-skalovane-anom} 
\end{figure*}

To investigate the observed deviations from linear dependences in more detail, we visualize the total and nonlinear contributions to dependence patterns in Figure~\ref{fig:lokalizace-skalovane-anom}. 
At first inspection one can see several well defined areas of relatively high nonlinear contribution to dependence patterns. These are in particular an extensive ring around the Anctarctica within the Southern Ocean, a few locations close to the North Pole (Barentz Sea, Bering Sea, Baffin Bay, Greenland Sea) and areas in Brazil and Southwest Asia.

\begin{figure*}
\includegraphics[width=1.0\textwidth]{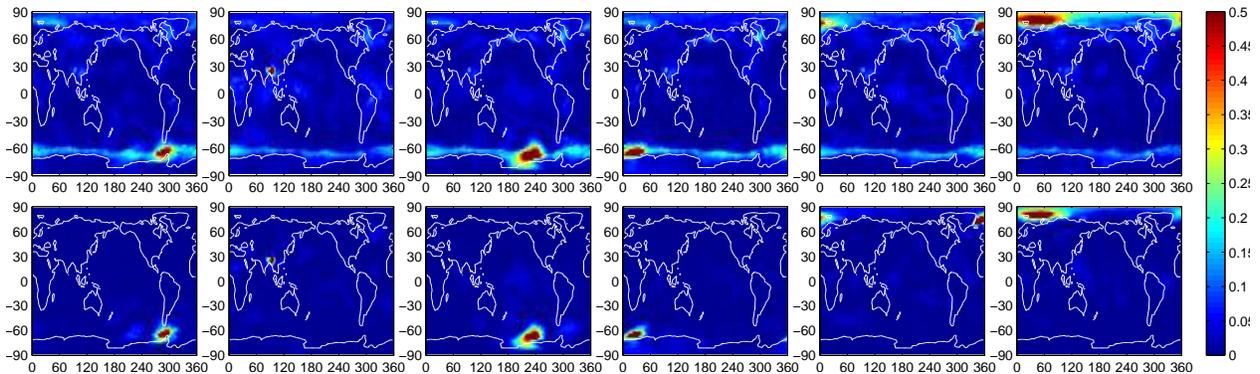}
\caption{Dependence patterns of six locations with high relative nonlinear dependence contribution. Top: total mutual information. Bottom: linear mutual information}
\label{fig:6FCpatterns} 
\end{figure*}

To understand the nonlinear dependence pattern in more detail, we select the locations with the highest relative nonlinear dependences and visualize both their linear and nonlinear dependence pattern with respect to all other locations, see Figure~\ref{fig:6FCpatterns}. For most of these areas, the nonlinear dependences are not generally stronger, but rather include additional distant locations, in contrast with the mostly local character of linear dependence patterns. 

\begin{figure}
\includegraphics[width=\columnwidth]{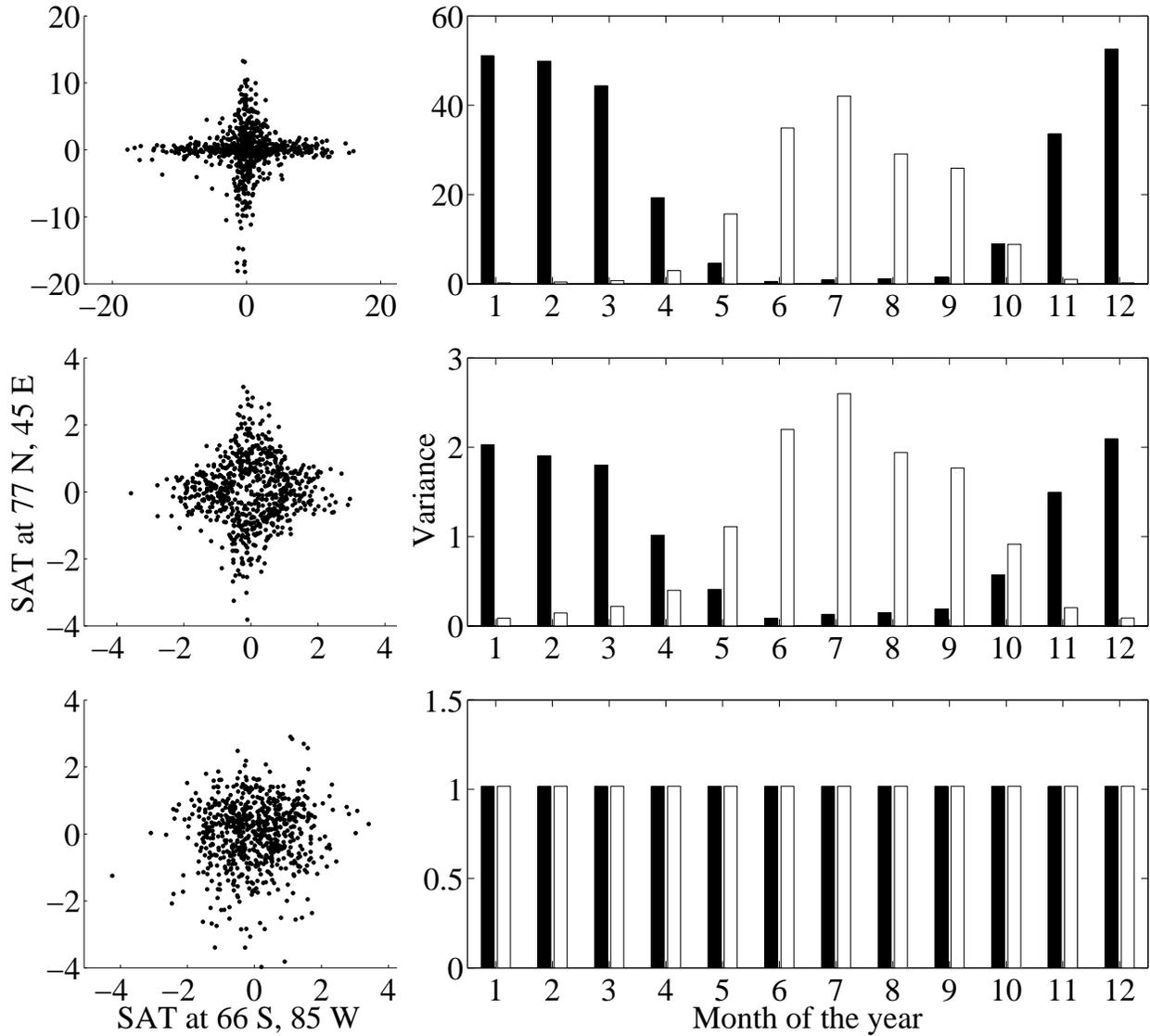}
\caption{Example of apparent nonlinear coupling between remote areas. The apparent nonlinearity is attributable to yearly cyclicity in variance, see text for details. Top: original data anomalies, middle: univariately normalized anomalies, bottom: monthly variance normalized anomalies. Left column: scatterplots of time series values, right column: variances of data for each month (black: 77 N, 45 E; white: 66 S, 85 W).}
\label{fig:Figure_scatter} 
\end{figure}

This might suggest the existence of long-range interactions or ``teleconnections'' of highly or predominantly non-linear character, as discussed e.g. in~\citep{Hsieh2006}. 
To elucidate the nature of these long-range connections we inspected the bivariate distributions and time series of the variables.  A representative example is shown in Figure~\ref{fig:Figure_scatter}. The shape of the bivariate distribution together with close inspection of the time series suggests that the non-Gaussianity might be related to seasonal variability in variance of the signal, which further differs between the two locations. 

In this particular case, the variability at the first location is the highest in December to February, when it is at its lowest at the second location and vice versa in July. Thus, the information shared by these time series would be explainable just by the seasonal differences of dynamics and ultimately just by variation in local solar influx. 

\begin{figure*}
\includegraphics[width=\textwidth]{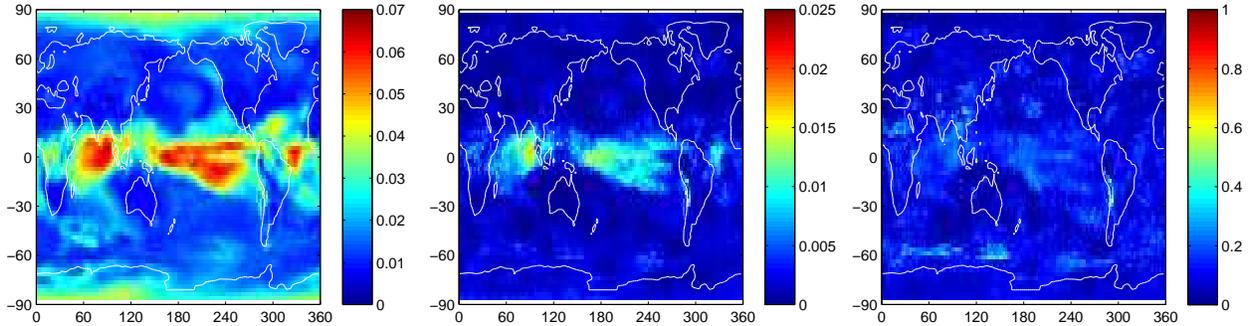}
\caption{Average contribution of nonlinear dependences in variance-normalized data.  Left: average mutual information for each location. Middle: average nonlinear contribution to mutual information $I_E$. Right: average nonlinear contribution relative to total mutual information ($I_E/I$).}
\label{fig:lokalizace-skalovane-varnorm} 
\end{figure*}

To test this hypotheses and control for this source of bias, we re-analyze the data after normalizing the seasonal variance, as described in Section~\ref{sec:mat}.
After this additional preprocessing step, there was a marked decrease in detected  pairs of locations with statistically significant  nonlinear contribution to temperature dependence ($8.6\%$ at the $p=0.05$ level), however, this is still more than expected by chance. The localization of nodes with the strongest (non)linear dependences is shown in Figure~\ref{fig:lokalizace-skalovane-varnorm}. 

By comparison to Figure~\ref{fig:lokalizace-skalovane-anom} we can see that the strongest contributors to apparent nonlinear dependences have been mitigated by this data cleaning step. The maxima of the relative nonlinear contributions are now much lower and are located almost purely in the equatorial ocean regions.   

\begin{figure}
\begin{center}
\includegraphics[width=\columnwidth]{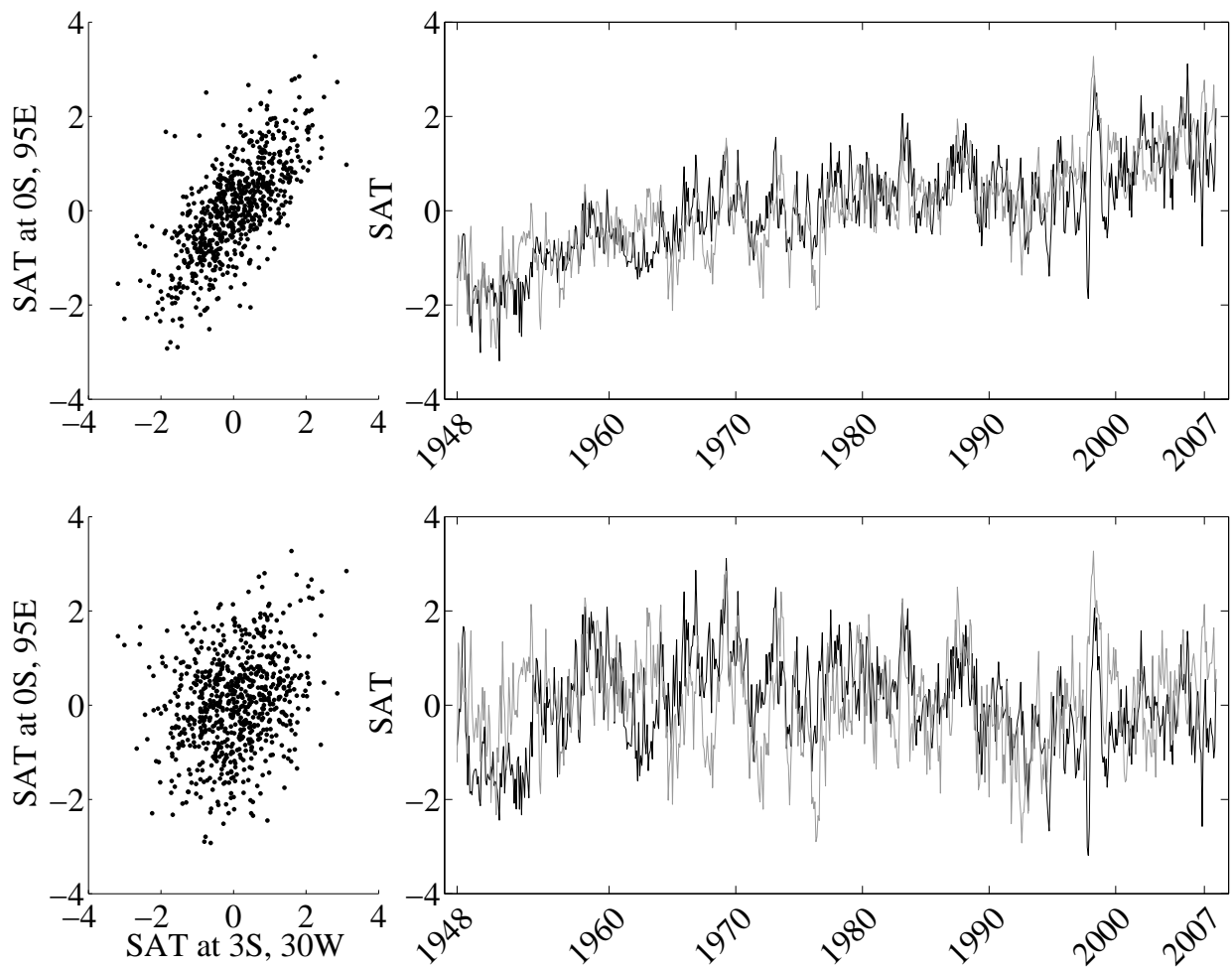}
\caption{Example of apparent nonlinear coupling between remote areas. The apparent (nonlinear) coupling is attributable to slow trend in the data, see text for details. Top: monthly variance normalized anomalies, bottom: after removal of linear trend. Left column: scatterplots of time series values, right column: time series (black: 0 N, 95 E; gray: 3 N, 30 W)}
\label{fig:Figure_scatter_varnorm} 
\end{center}
\end{figure}

As previously, we investigate the form of a nonlinear dependence pattern related to the strongest source of nonlinearity in thus preprocessed data. 
This is shown in Figure~\ref{fig:Figure_scatter_varnorm}, showing typical example bivariate distributions and time series. 
The source of the observed non-Gaussianity of bivariate dependence can be commonly tracked down to an apparent non-stationarity of the time series. In particular in the example shown in Figure~\ref{fig:Figure_scatter_varnorm} there is a strong (almost linear) trend, which might be of interest for other reasons, but can be considered as spurious with respect to detection of climate interactions on the time scale considered.

This motivates additional detrending of the time series followed by yet another replication of the analysis. 

\begin{figure*}
\includegraphics[width=\textwidth]{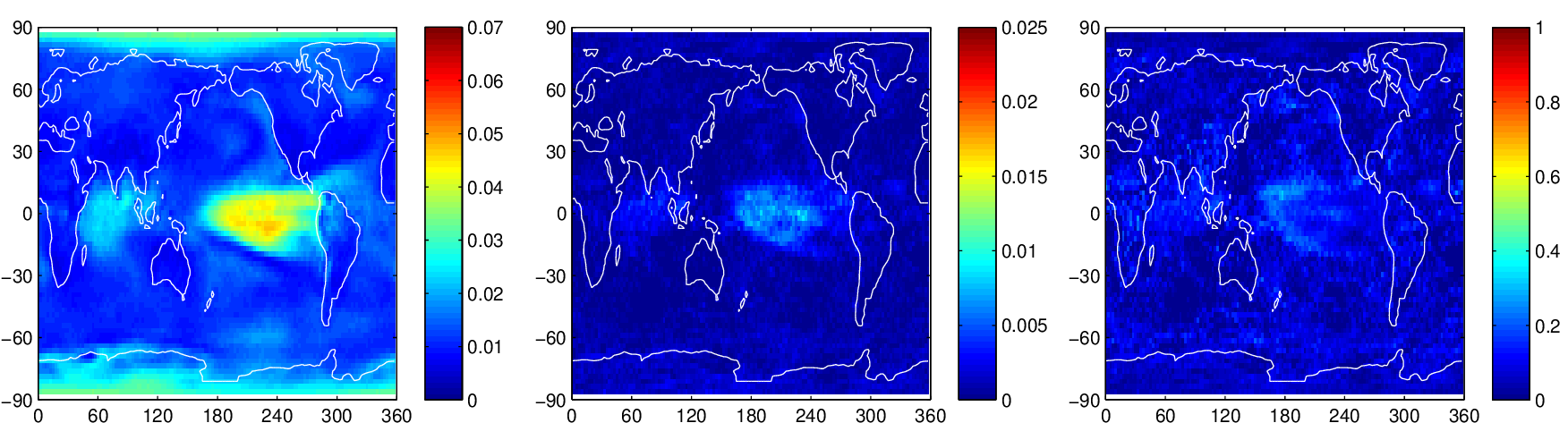}
\caption{Average contribution of nonlinear dependences in detrended variance-normalized data. Left: average mutual information for each location. Middle: average nonlinear contribution to mutual information $I_E=I-I_{surr}$. Right: relative average nonlinear contribution to mutual information ($I_E/I$).}
\label{fig:lokalizace-skalovane-varnorm-detrend} 
\end{figure*}

The results suggest that the trends are indeed responsible for a major part of the yet remaining apparent non-linearity. In particular, the amount of detected pairs of locations with statistically significant nonlinear contribution to temperature dependence goes further down to $6.5\%$ at the $p=0.05$ level.  Similarly, the average nonlinear contribution to mutual information is further substantially reduced, see Figure~\ref{fig:lokalizace-skalovane-varnorm-detrend}.

\subsection{ERA dataset}

The analysis was replicated for the ERA dataset. 

\begin{figure}
\includegraphics[width=\columnwidth]{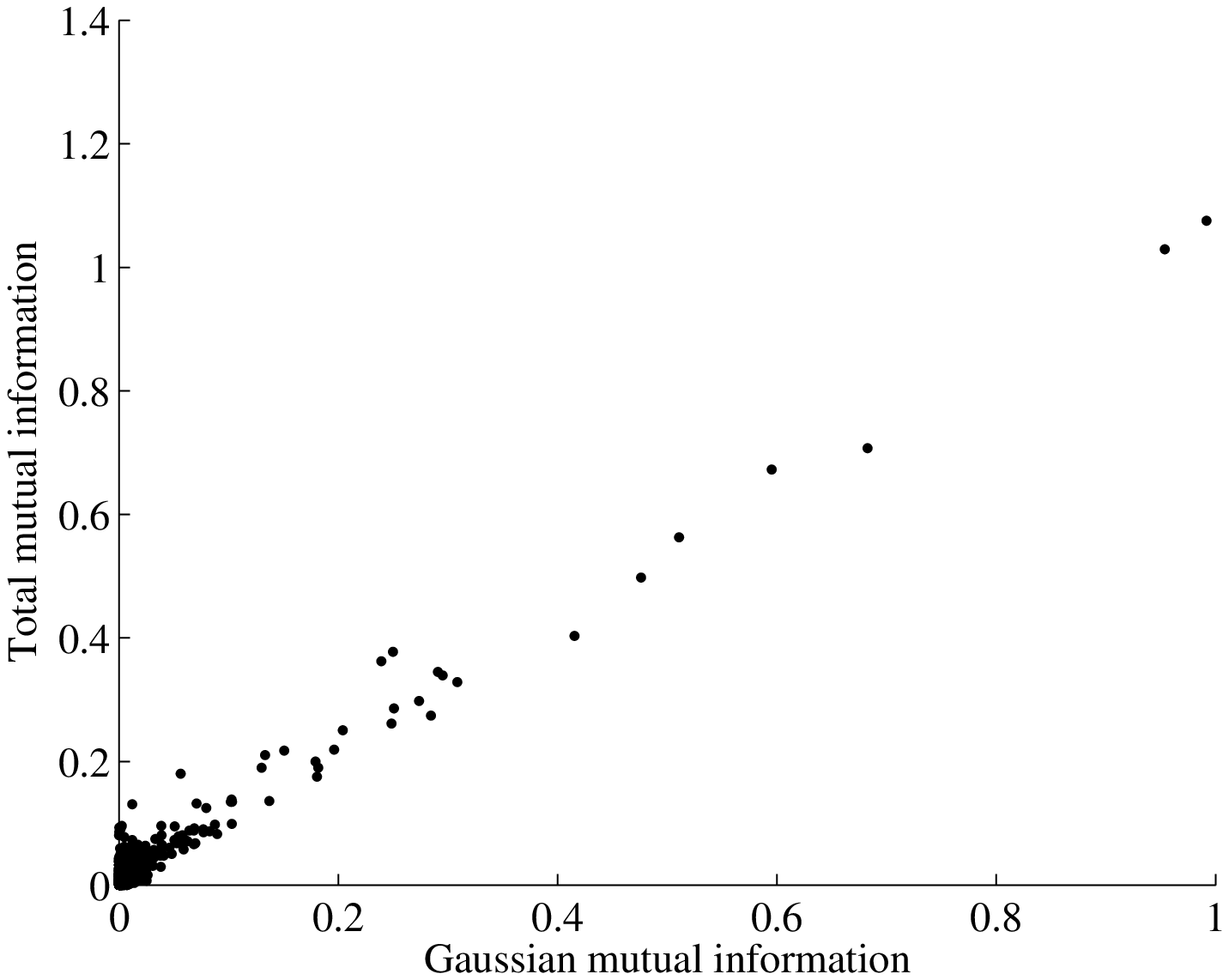} 
\caption{Relation between total and linear mutual information estimates for the ERA dataset. Each blue dot corresponds to one time series pair. For visualization purposes, a random selection of 1000 (out of over 5 million) time series pairs shown.}
\label{fig:FigureMIvsMI_ERA} 
\end{figure}

\begin{figure*}
\includegraphics[width=\textwidth]{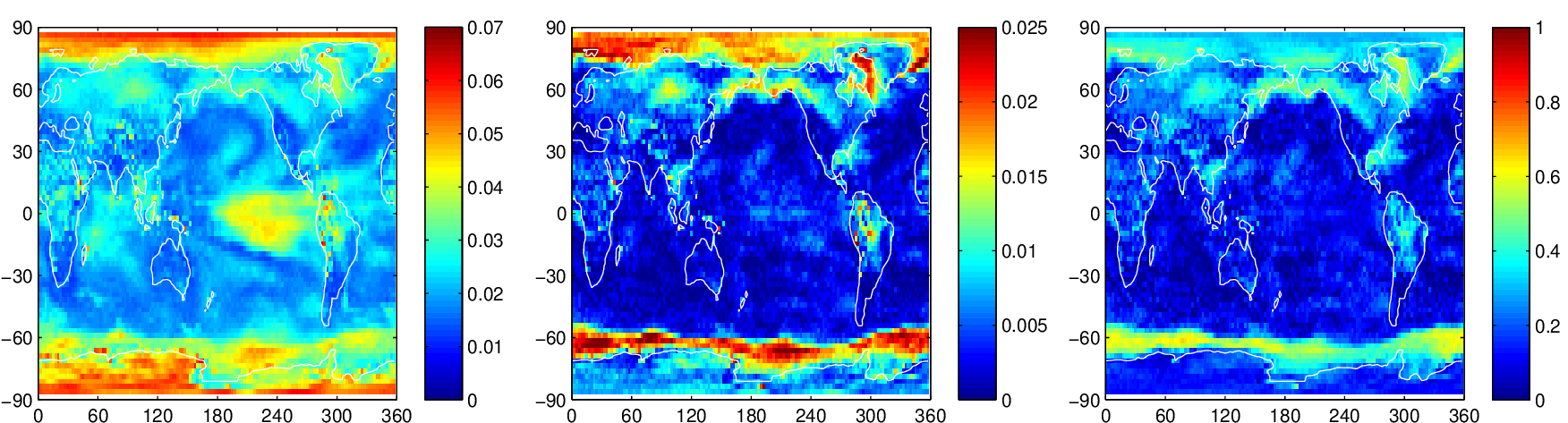}
\caption{Average contribution of nonlinear dependences for the ERA dataset. Left: average mutual information for each location. Middle: average nonlinear contribution to mutual information $I_E$. Right: average nonlinear contribution relative to total mutual information ($I_E/I$).}
\label{fig:lokalizace-skalovane-anom_ERA} 
\end{figure*}

\begin{figure*}
\includegraphics[width=\textwidth]{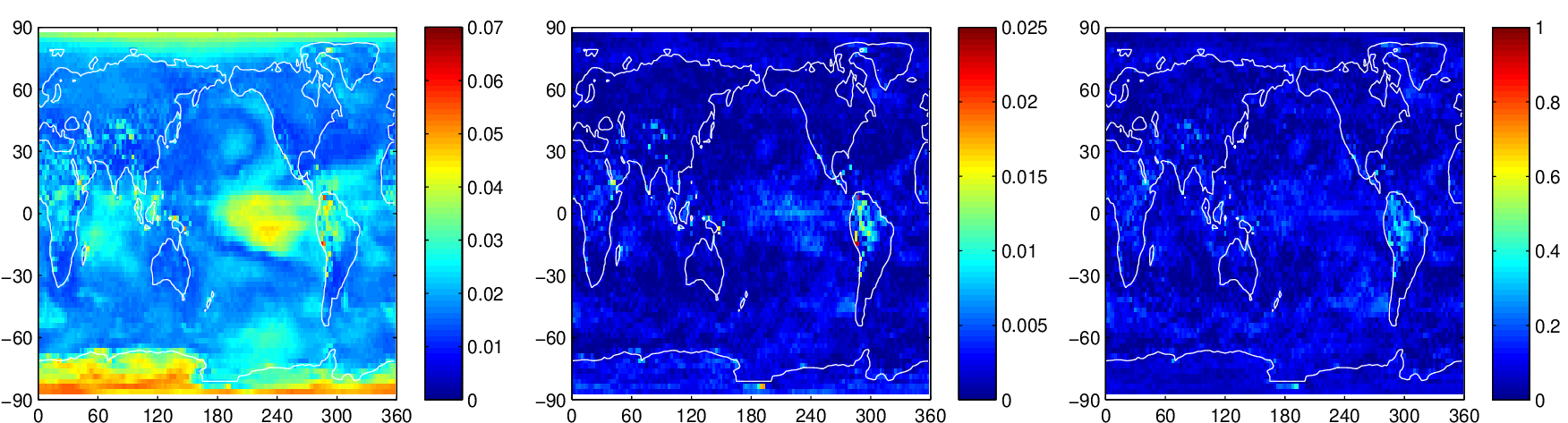}
\caption{Average contribution of nonlinear dependences in variance-normalized data for the ERA dataset.  Left: average mutual information for each location. Middle: average nonlinear contribution to mutual information $I_E$. Right: average nonlinear contribution relative to total mutual information ($I_E/I$).}
\label{fig:lokalizace-skalovane-varnorm_ERA} 
\end{figure*}

\begin{figure*}
\includegraphics[width=\textwidth]{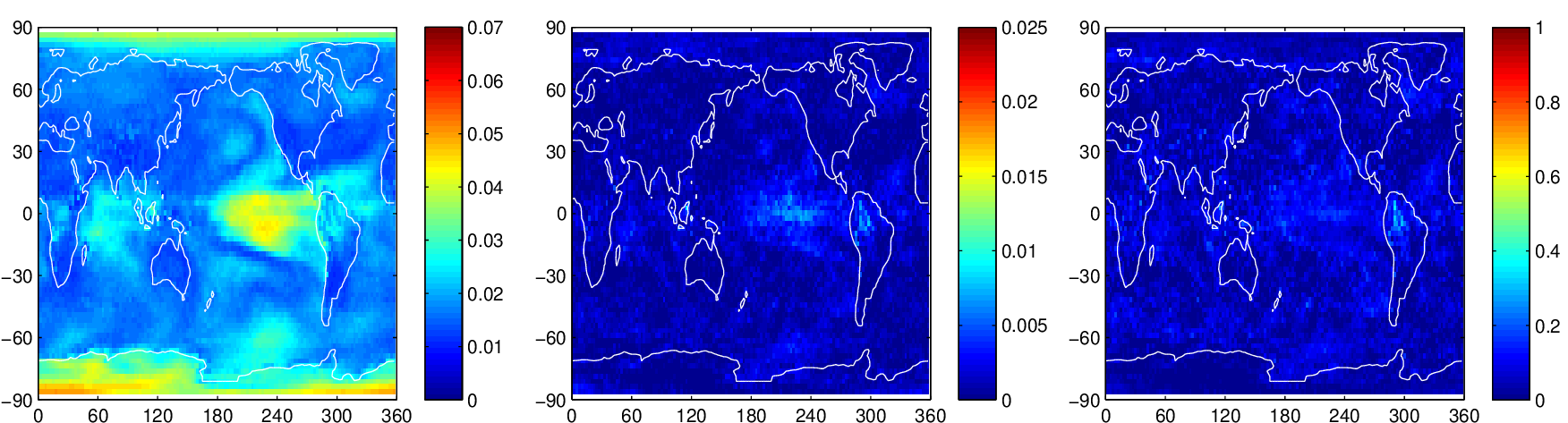}
\caption{Average contribution of nonlinear dependences in detrended variance-normalized data for the ERA dataset. Left: average mutual information for each location. Middle: average nonlinear contribution to mutual information $I_E=I-I_{surr}$. Right: relative average nonlinear contribution to mutual information ($I_E/I$).}
\label{fig:lokalizace-skalovane-varnorm-detrend_ERA} 
\end{figure*}

The general strength and distribution of nonlinear dependence within the ERA dataset is very similar to the NCEP/NCAR, see Figures~\ref{fig:FigureMIvsMI_ERA}, \ref{fig:lokalizace-skalovane-anom_ERA}, \ref{fig:lokalizace-skalovane-varnorm_ERA} and ~\ref{fig:lokalizace-skalovane-varnorm-detrend_ERA}.
The fraction of pairs of nodes with significant nongaussianity were at the $p=0.05$ level is $16.3$, $7.0$ and $6.1$ percent in the original anomalies, variance normalized data and detrended variance normalized data respectively, suggesting just a slightly weaker contribution of the nonlinearity in the ERA dataset, and clearly not much more than expected by chance for linear data in the last case. 

\section{Discussion}
\label{sec:dis}

In the previous sections we have outlined an approach for a detailed multi-step analysis of the relevance of nonlinearity in the dependence structure of climate time series and the results for monthly SAT reanalysis data. The overall picture is in general that of negligible nonlinearities in the data; the most substantial apparent `nonlinearities' are attributable to nonstationarity effects. 

Therefore, the tentative suggestion with respect to choice of dependence measure for this type of data would be to use a linear measure (Pearson correlation coefficient), potentially after removing data nonstationarity by preprocessing, as some of the nonstationarities may affect also Pearson correlation estimates, albeit differently than the nonlinear mutual information.

An obvious question is that of generalizability of the specific findings. Explicitly, we have confirmed similar results in the ERA reanalysis SAT data.

Also in daily NCEP/NCAR data we have observed relatively negligible nonlinearity, after removing nonstationarities in variance as well as trends (results not shown).    


While we have only showed results for the most commonly used grid with fixed angular resolution $2.5^{\circ}$, based on inspection of the spatiotemporal structure of the data we conjecture that analogous results would be observed for other resolutions as well as area-corrected grids as used e.g. by~\citep{Donges2011}.

The present results extend those of ~\citet{Palus1994}
who tested for possible nonlinearity in the dynamics of the station
(Prague-Klementinum) SAT time series and found that
the dependence between the SAT time
series $\{x(t)\}$ and its lagged twin $\{x(t + \tau)\}$ was
well-explained by a linear stochastic process. This result about
a linear character of the temporal evolution of SAT time series, as well
as the results of this study about relations between the reanalysis
SAT time series from different grid-points cannot be understood as
arguments for a linear character of atmospheric dynamics. These
results rather characterize properties of measurement or reanalysis
data at a particularly coarse level of resolution when the data reflecting
a spatially and temporally averaged mixture of dynamical processes
on a wide range of spatial and temporal scales are considered.
For instance,
a closer look on the dynamics on
specific temporal scales in temperature and other meteorological
data has led to identification of oscillatory phenomena
with nonlinear behavior, exhibiting phase synchronization~
\citep{Palus2004,Palus2006,Palus2009, Palus2011a, Feliks2010}.

Also for other variables with vastly different dynamics we can expect substantial nonlinearity in bivariate dependencies, especially if the measurement/model has sufficient spatial and temporal resolution. Conceptually, similar analysis to that presented is warranted to be carried out before the decision for use of linear or nonlinear dependence measure for each substantially different dataset. The work on preparation of a semi-automated tool for such analysis is undergoing.


Despite the relative sparsity of the substantially nonlinear dependence patterns, even after the two additional preprocessing steps we have observed more than the expected proportion of significantly nonlinear dependences ($6\%$ instead of expected $5\%$). Given the number of tests carried out (several million of location-pairs), this small deviation likely consists a globally significant deviation from multivariate Gaussianity, although the intricate interdependence of the pair tests themselves makes the estimation of the exact p-value for a global linearity hypothesis technically very difficult.

From practical point of view, it may be argued that the statistical determination of the above-random presence of apparent nonlinearity should not play a key role in method choice. Firstly, the more detailed quantitative analysis has already shown that even where present, the effect is relatively weak. Secondly, even though for some region pairs there may be a statistical indication of nonlinear dependence, it is realistic to suspect (given the results for the raw data, see Figures~\ref{fig:Figure_scatter} and \ref{fig:Figure_scatter_varnorm}) that this apparent nonlinearity may be due to as yet not discussed type of nonstationarity, and visual inspection of the data is required. Indeed, several further spurious nonlinearities have been detected due to various uncorrected problems within the reanalysis data (some of which corresponded to known problems as described in~\citep{Kistler2001}).

\begin{figure}
\begin{center}
\includegraphics[width=\columnwidth]{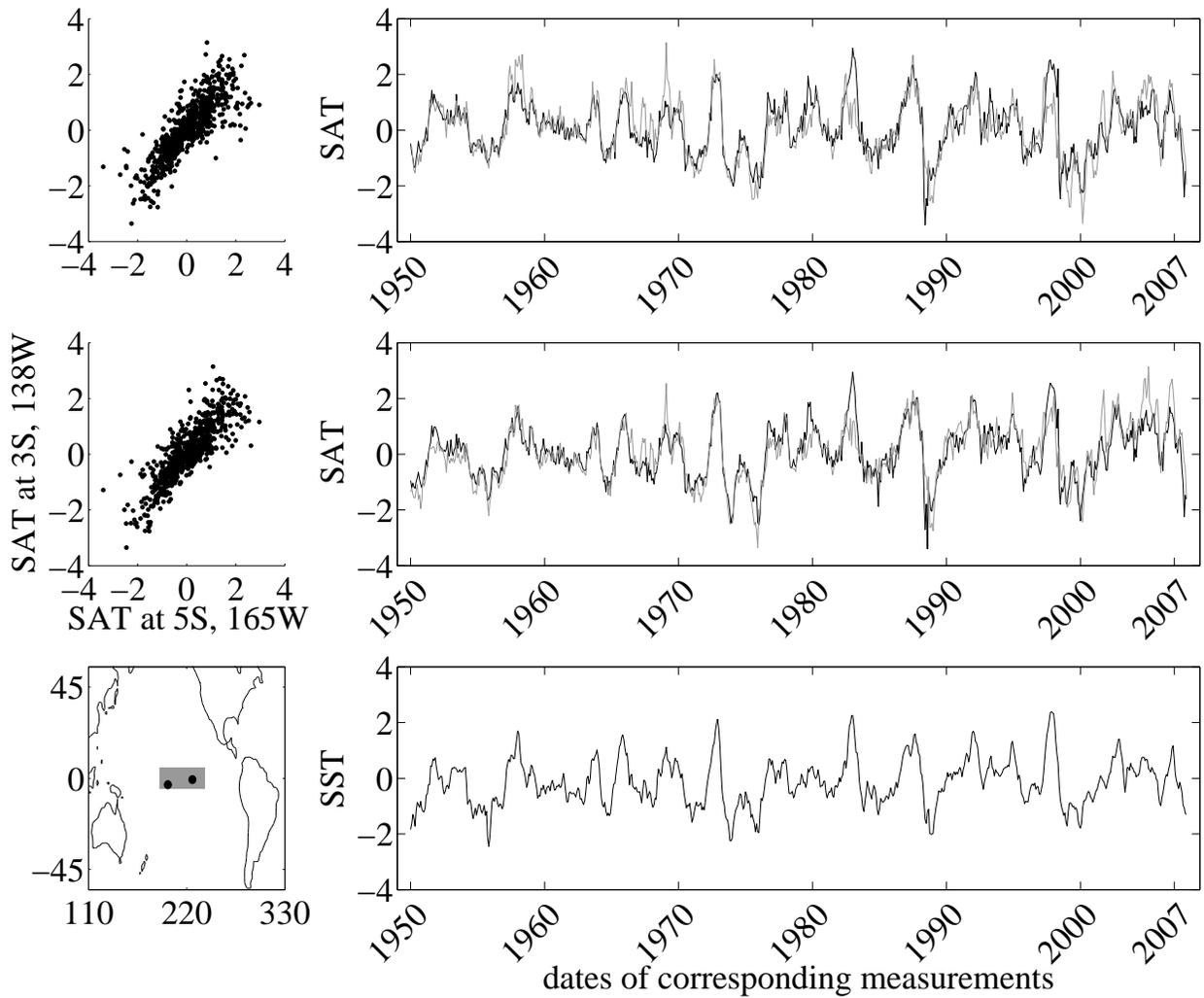}
\caption{Example of one of the couplings with substantial residual nonlinear contribution even after removal of several sources of spurious nonlinearity: between two locations in the Pacific Ocean (the dependence pattern still substantially deviates from bivariate Gaussianity, with the estimated total and linear mutual information being 0.73 and 0.45 nats respectively; see bottom left for approximate location of the points). The apparent (nonlinear) coupling may be attributable to nonlinearity of ENSO dynamics. Top: monthly SAT anomalies after seasonal variance normalization and linear trend removal (left: scatterplots of time series values, right: time series - black: 5S, 165W; gray: 3S, 138W), middle: monthly SAT anomalies without seasonal variance normalization and linear trend removal, bottom left: depiction of the respective locations and the SST averaging region for obtaining the El Nino 3.4 index, bottom right: El Nino 3.4 index time series.  See text for more details.} 
\label{fig:NonlinearRemainder} 
\end{center}
\end{figure}

This leads to the consideration of the strongest nonlinear pattern observed in the current data even after the two additional preprocessing steps on top of those carried out in~\citep{Donges2009}. As shown in Figure~\ref{fig:lokalizace-skalovane-varnorm-detrend}, ocean areas particularly in the tropical Pacific bear still a slightly elevated non-Gaussian contribution to dependence patterns. An example of such dependence pattern and related time series is shown in Figure~\ref{fig:NonlinearRemainder}. Note that the area of strongest residual non-Gaussianity roughly corresponds to regions implicated in ENSO dynamics; for visual comparison we plot the area and time series for the El Nino 3.4 index (the Nino 3.4 region is bounded by $120^\circ$W-$170^\circ$W and $5^\circ$S-$5^\circ$N~\citep{Trenberth1997}). The NINO3.4 SST index data were downloaded from from NOAA/NWS Climate Prediction Center, (http://www.cpc.ncep.noaa.gov, accessed November 15th, 2012). We hypothesize that the observed pattern reflects the nonlinear character of ENSO dynamics observed e.g. by ~\citet{An2004, Hannachi2003,Boucharel2011}.

\section{Conclusions}
\label{sec:con}

Quantification of dependences between variables is a common task within the study of climate or other complex systems.  
The choice of dependence measures is commonly based on some theoretical assumptions about the underlying system. In case of climatic time series, this may lead to the choice of nonlinear methods due to the nonlinear nature of the underlying physical processes. However, for various reasons including spatiotemporal sampling or averaging, measured or modeled data might be in fact well captured by linear measures, with their nonlinear counterparts potentially reducing sensitivity or introducing bias.


We have presented a multi-step approach that allows the detailed assessment of the nonlinear contribution to dependence patterns in a dataset, including not only statistical testing, but also quantification, localization and analysis of sources of this contribution. This approach can provide rationale for the decision regarding the choice of suitable dependence measure for given type of data, as well as direct the analyst attention to hidden crucial properties of the dataset.
Importantly, the presented approach is quite general. It is transferable to other geoscientific datasets, as well as to other disciplines such as neuroscience, see e.g. ~\citep{Hlinka2011Neuroimage} for an earlier application of a related approach and ~\citep{Hartman2011} for an example focused to an assessment of the nonlinearity effects on graph-theoretical network characteristics. Let us note that the construction and interpretation of graphs representing climate networks poses further challenges~\citep{Palus2011, Hlinka2012}.


For monthly SAT in the NCEP/NCAR dataset and similar data, the analysis suggests that the use of linear dependence methods is generally sufficient, potentially after the treatment of described nonstationarity sources.


This has shown that the quantitative study of amount of nonlinearity in the data provides as a side product valuable hints about potentially hidden specific  properties of the data, that may otherwise go unnoticed and bias the results, were just a single method used without detailed analysis.

\section*{Acknowledgements}
This study is supported by the Czech Science Foundation, Project No. P103/11/J068.

\clearpage 


\bibliographystyle{spbasic}      

\begin{thebibliography}{32}
\providecommand{\natexlab}[1]{#1}
\providecommand{\url}[1]{{#1}}
\providecommand{\urlprefix}{URL }
\expandafter\ifx\csname urlstyle\endcsname\relax
  \providecommand{\doi}[1]{DOI~\discretionary{}{}{}#1}\else
  \providecommand{\doi}{DOI~\discretionary{}{}{}\begingroup
  \urlstyle{rm}\Url}\fi
\providecommand{\eprint}[2][]{\url{#2}}

\bibitem[{An and Jin(2004)}]{An2004}
An S, Jin F (2004) Nonlinearity and asymmetry of {ENSO}. Journal of Climate
  17(12):2399--2412

\bibitem[{Boucharel et~al(2011)Boucharel, Dewitte, du~Penhoat, Garel, Yeh, and
  Kug}]{Boucharel2011}
Boucharel J, Dewitte B, du~Penhoat Y, Garel B, Yeh SW, Kug JS (2011) {ENSO}
  nonlinearity in a warming climate. Climate Dynamics 37(9-10):2045--2065

\bibitem[{Dee et~al(2011)Dee, Uppala, Simmons, Berrisford, Poli, Kobayashi,
  Andrae, Balmaseda, Balsamo, Bauer, Bechtold, Beljaars, van~de Berg, Bidlot,
  Bormann, Delsol, Dragani, Fuentes, Geer, Haimberger, Healy, Hersbach, Holm,
  Isaksen, Kallberg, Koehler, Matricardi, McNally, Monge-Sanz, Morcrette, Park,
  Peubey, de~Rosnay, Tavolato, Thepaut, and Vitart}]{Dee2011}
Dee DP, Uppala SM, Simmons AJ, Berrisford P, Poli P, Kobayashi S, Andrae U,
  Balmaseda MA, Balsamo G, Bauer P, Bechtold P, Beljaars ACM, van~de Berg L,
  Bidlot J, Bormann N, Delsol C, Dragani R, Fuentes M, Geer AJ, Haimberger L,
  Healy SB, Hersbach H, Holm EV, Isaksen L, Kallberg P, Koehler M, Matricardi
  M, McNally AP, Monge-Sanz BM, Morcrette JJ, Park BK, Peubey C, de~Rosnay P,
  Tavolato C, Thepaut JN, Vitart F (2011) The {ERA-Interim} reanalysis:
  configuration and performance of the data assimilation system. Quarterly
  Journal of the Royal Meteorological Society 137(656, Part a):553--597

\bibitem[{Diks and Mudelsee(2000)}]{Diks2000}
Diks C, Mudelsee M (2000) Redundancies in the earth's climatological time
  series. Physics Letters A 275(5-6):407--414

\bibitem[{Donges et~al(2009)Donges, Zou, Marwan, and Kurths}]{Donges2009}
Donges JF, Zou Y, Marwan N, Kurths J (2009) The backbone of the climate
  network. EPL 87(4):48,007

\bibitem[{Donges et~al(2011)Donges, Schultz, Marwan, Zou, and
  Kurths}]{Donges2011}
Donges JF, Schultz HCH, Marwan N, Zou Y, Kurths J (2011) Investigating the
  topology of interacting networks theory and application to coupled climate
  subnetworks. European Physical Journal B 84(4):635--651

\bibitem[{Feliks et~al(2010)Feliks, Ghil, and Robertson}]{Feliks2010}
Feliks Y, Ghil M, Robertson AW (2010) Oscillatory climate modes in the eastern
  mediterranean and their synchronization with the north atlantic oscillation.
  Journal of Climate 23(15):4060--4079

\bibitem[{Hannachi et~al(2003)Hannachi, Stephenson, and Sperber}]{Hannachi2003}
Hannachi A, Stephenson D, Sperber K (2003) Probability-based methods for
  quantifying nonlinearity in the {ENSO}. Climate Dynamics 20(2-3):241--256

\bibitem[{Hannachi et~al(2007)Hannachi, Jolliffe, and
  Stephenson}]{Hannachi2007}
Hannachi A, Jolliffe IT, Stephenson DB (2007) Empirical orthogonal functions
  and related techniques in atmospheric science: A review. International
  Journal Of Climatology 27(9):1119--1152

\bibitem[{Hartman et~al(2011)Hartman, Hlinka, Palu\v{s}, Mantini, and
  Corbetta}]{Hartman2011}
Hartman D, Hlinka J, Palu\v{s} M, Mantini D, Corbetta M (2011) The role of
  nonlinearity in computing graph-theoretical properties of resting-state
  functional magnetic resonance imaging brain networks. Chaos 21(1)

\bibitem[{Hlinka et~al(2011)Hlinka, Palus, Vejmelka, Mantini, and
  Corbetta}]{Hlinka2011Neuroimage}
Hlinka J, Palus M, Vejmelka M, Mantini D, Corbetta M (2011) {Functional
  connectivity in resting-state fMRI: Is linear correlation sufficient?}
  NeuroImage 54:2218--2225

\bibitem[{Hlinka et~al(2012)Hlinka, Hartman, and Palus}]{Hlinka2012}
Hlinka J, Hartman D, Palus M (2012) Small-world topology of functional
  connectivity in randomly connected dynamical systems. Chaos 22(3)

\bibitem[{Hsieh et~al(2006)Hsieh, Wu, and Shabbar}]{Hsieh2006}
Hsieh W, Wu A, Shabbar A (2006) Nonlinear atmospheric teleconnections.
  Geophysical Research Letters 33(7)

\bibitem[{Kalnay et~al(1996)Kalnay, Kanamitsu, Kistler, Collins, Deaven,
  Gandin, Iredell, Saha, White, Woollen, Zhu, Chelliah, Ebisuzaki, Higgins,
  Janowiak, Mo, Ropelewski, Wang, Leetmaa, Reynolds, Jenne, and
  Joseph}]{Kalnay1996}
Kalnay E, Kanamitsu M, Kistler R, Collins W, Deaven D, Gandin L, Iredell M,
  Saha S, White G, Woollen J, Zhu Y, Chelliah M, Ebisuzaki W, Higgins W,
  Janowiak J, Mo K, Ropelewski C, Wang J, Leetmaa A, Reynolds R, Jenne R,
  Joseph D (1996) The {NCEP/NCAR} 40-year reanalysis project. Bulletin of the
  American Meteorological Society 77(3):437--471

\bibitem[{Kendall(1938)}]{Kendall1938}
Kendall M (1938) A new measure of rank correlation. Biometrika 30(Part
  1/2):81--93

\bibitem[{Kistler et~al(2001)Kistler, Kalnay, Collins, Saha, White, Woollen,
  Chelliah, Ebisuzaki, Kanamitsu, Kousky, van~den Dool, Jenne, and
  Fiorino}]{Kistler2001}
Kistler R, Kalnay E, Collins W, Saha S, White G, Woollen J, Chelliah M,
  Ebisuzaki W, Kanamitsu M, Kousky V, van~den Dool H, Jenne R, Fiorino M (2001)
  The {NCEP-NCAR} 50-year reanalysis: Monthly means {CD-ROM} and documentation.
  Bulletin of the American Meteorological Society 82(2):247--267

\bibitem[{Palus(1997)}]{Palus1997}
Palus M (1997) Detecting phase synchronization in noisy systems. Physics
  Letters A 235(4):341--351

\bibitem[{Palus and Novotna(1994)}]{Palus1994}
Palus M, Novotna D (1994) Testing for nonlinearity in weather records. Physics
  Letters A 193(1):67--74

\bibitem[{Palus and Novotna(2004)}]{Palus2004}
Palus M, Novotna D (2004) Enhanced monte carlo singular system analysis and
  detection of period 7.8 years oscillatory modes in the monthly {NAO} index
  and temperature records. Nonlinear Processes in Geophysics 11(5-6):721--729

\bibitem[{Palus and Novotna(2006)}]{Palus2006}
Palus M, Novotna D (2006) Quasi-biennial oscillations extracted from the
  monthly nao index and temperature records are phase-synchronized. Nonlinear
  Processes in Geophysics 13(3):287--296

\bibitem[{Palus and Novotna(2009)}]{Palus2009}
Palus M, Novotna D (2009) Phase-coherent oscillatory modes in solar and
  geomagnetic activity and climate variability. Journal of Atmospheric and
  Solar-terrestrial Physics 71(8-9):923--930

\bibitem[{Palus and Novotna(2011)}]{Palus2011a}
Palus M, Novotna D (2011) Northern hemisphere patterns of phase coherence
  between solar/geomagnetic activity and ncep/ncar and era40 near-surface air
  temperature in period 7-8 years oscillatory modes. Nonlinear Processes in
  Geophysics 18(2):251--260

\bibitem[{Palus and Vejmelka({2007})}]{Palus2007a}
Palus M, Vejmelka M ({2007}) {Directionality of coupling from bivariate time
  series: How to avoid false causalities and missed connections}. {Physical
  Review E} {75}({5, Part 2}):056,211

\bibitem[{Palus et~al(1993)Palus, Albrecht, and Dvorak}]{Palus1993}
Palus M, Albrecht V, Dvorak I (1993) Information theoretic test for
  nonlinearity in time series. Physics Letters A 175({3-4}):203--209

\bibitem[{Palus et~al(2011)Palus, Hartman, Hlinka, and Vejmelka}]{Palus2011}
Palus M, Hartman D, Hlinka J, Vejmelka M (2011) Discerning connectivity from
  dynamics in climate networks. Nonlinear Processes in Geophysics
  18(5):751--763

\bibitem[{Prichard and Theiler(1994)}]{Prichard1994}
Prichard D, Theiler J (1994) Generating surrogate data for time series with
  several simultaneously measured variables. Physical Review Letters 73(7):951

\bibitem[{Shannon(1948)}]{Shannon1948}
Shannon C (1948) A mathematical theory of communication. Bell System Technical
  Journal 27(3):379--423

\bibitem[{Spearman(1904)}]{Spearman1904}
Spearman C (1904) The proof and measurement of association between two things.
  American Journal of Psychology 15:72--101

\bibitem[{Trenberth(1997)}]{Trenberth1997}
Trenberth K (1997) The definition of el nino. Bulletin of the American
  Meteorological Society 78(12):2771--2777

\bibitem[{Tsonis and Roebber(2004)}]{Tsonis2004}
Tsonis A, Roebber P (2004) The architecture of the climate network. Physica A
  333:497--504

\bibitem[{Tsonis et~al(2011)Tsonis, Wang, Swanson, Rodrigues, and
  Costa}]{Tsonis2011}
Tsonis AA, Wang G, Swanson KL, Rodrigues FA, Costa LdF (2011) Community
  structure and dynamics in climate networks. Climate Dynamics 37(5-6):933--940

\bibitem[{Uppala et~al(2005)Uppala, Kallberg, Simmons, Andrae, Bechtold,
  Fiorino, Gibson, Haseler, Hernandez, Kelly, Li, Onogi, Saarinen, Sokka,
  Allan, Andersson, Arpe, Balmaseda, Beljaars, Van De~Berg, Bidlot, Bormann,
  Caires, Chevallier, Dethof, Dragosavac, Fisher, Fuentes, Hagemann, Holm,
  Hoskins, Isaksen, Janssen, Jenne, McNally, Mahfouf, Morcrette, Rayner,
  Saunders, Simon, Sterl, Trenberth, Untch, Vasiljevic, Viterbo, and
  Woollen}]{Uppala2005}
Uppala S, Kallberg P, Simmons A, Andrae U, Bechtold V, Fiorino M, Gibson J,
  Haseler J, Hernandez A, Kelly G, Li X, Onogi K, Saarinen S, Sokka N, Allan R,
  Andersson E, Arpe K, Balmaseda M, Beljaars A, Van De~Berg L, Bidlot J,
  Bormann N, Caires S, Chevallier F, Dethof A, Dragosavac M, Fisher M, Fuentes
  M, Hagemann S, Holm E, Hoskins B, Isaksen L, Janssen P, Jenne R, McNally A,
  Mahfouf J, Morcrette J, Rayner N, Saunders R, Simon P, Sterl A, Trenberth K,
  Untch A, Vasiljevic D, Viterbo P, Woollen J (2005) The {ERA}-40 re-analysis.
  Quarterly Journal of the Royal Meteorological Society 131(612, Part
  b):2961--3012

\end{thebibliography}


\end{document}